\documentclass[11pt,a4paper,english]{article}
\usepackage{times}
\usepackage[pdftex]{graphicx}
\usepackage{amsfonts}
\usepackage{setspace}
\usepackage{tabularx}
\usepackage{amssymb}
\usepackage{amsthm}
\usepackage{bm}
\usepackage{graphicx}
\usepackage{amsmath,scalerel}%
\usepackage{subcaption}
\usepackage{color}
\usepackage{hyperref}
\usepackage{eurosym}
\usepackage{eucal}
\usepackage{float}
\usepackage{multirow}
\usepackage[T1]{fontenc}
\usepackage[round]{natbib}
\usepackage{epstopdf}

\begin{document}
	\baselineskip20pt
	\abovedisplayskip0.5cm
	\belowdisplayskip0.5cm
	\setlength{\topsep}{0.5cm}
	\title{Bayesian nonparametric copulas with tail dependence}
	\author{M.\ C.\ Aus\'in \footnote{Department of Statistics, Universidad Carlos III de Madrid, Spain, concepcion.ausin@uc3m.es}\qquad M. Kalli \footnote{Department of Mathematics, King's College London, UK, maria.kalli@kcl.ac.uk}}
	\date{}
	\maketitle
	%
	%
	\thispagestyle{empty}
	\vspace*{-1cm}
	\moveleft\hoffset\vbox{\hrule width\textwidth height 1pt}

\begin{abstract}
We introduce a novel bivariate copula model able to capture both the central and tail dependence of the joint probability distribution. Models that can capture the dependence structure within the joint tail have important implications in applications focusing on risk management (e.g macroeconomics and finance). We use a Bayesian nonparametric approach to introduce a random copula model based on infinite partitions of unity. We define a hierarchical prior over an infinite partition of the unit hypercube which has a stick-breaking representation leading to an infinite mixture of products of independent beta densities. Capitalising on the stick-breaking representation we introduce a Gibbs sampler to proceed to inference. For our empirical analysis we consider both simulated and real data (insurance claims and portfolio returns). We compare both our model's ability to capture tail dependence and its out-of-sample predictive performance to parametric and nonparametric competitive copula models. We show that in both simulated and real examples our model outperforms the competitive models.
\end{abstract}



\moveleft\hoffset\vbox{\hrule width\textwidth height 1pt}

%
%
\newpage
\setcounter{page}{1}
\baselineskip20pt
\renewcommand{\thefootnote}{\arabic{footnote}} 

\section{Introduction}

The most interesting problems in finance involve modelling multivariate data sets with complex dependence structures. Modelling the interactions between variables such as asset returns and capturing the tail dependence of their joint distribution is key in managing investment portfolio risk \citep[see][]{fort02, furm16, cort19}. Tail dependence refers to the clustering of extreme events, and in finance these are high severity risks (market crashes, wars, pandemics, e.t.c.) that can lead to heavy investment portfolio losses.  The importance of accounting for tail dependence in the presence of long-tailed claims variables for insurance risk management to mitigate potential losses is emphasised in \cite{hogg84} and \cite{free98}.  In portfolio management, \cite{embrechts97}, and \cite{emb02} point out that tail as well as central dependence structure of the joint financial asset distribution matters. This is due to the shared aims of minimising the probability of large losses and being able to re-allocate funds between assets in order to maximise the investors' expected utility.

Modelling the joint distribution of financial variables is not an easy task. A flexible approach is needed in order to capture both the tail and central dependence structure together with the slight asymmetry of the distribution. \cite{ledford97} highlight two issues when it comes to modelling the joint tail of an unknown multivariate distribution. One is modelling the tail of each marginal distribution and the other is modelling the dependence structure within the joint tail.  We focus on modelling the joint tail and asymmetry. Resorting to multivariate distributions of known parameter families such as the Gaussian or the student-t does not allow for the needed flexibility to account for these issues. Our approach to constructing the joint multivariate distribution builds on copula functions introduced by \cite{sklar59}. Sklar's theorem states that for a vector of random variables ${\bm y}= (y_1, y_2, \ldots, y_d)'$ with cumulative distribution function $F$ and $F_i$ the marginal distribution of $y_i$ for $i= 1,2, \ldots, d$, there exists a copula function $C \,:\, [0,1]^d\rightarrow [0,1]$ such that for all ${\bm y} =(y_1,y_2,\ldots, y_d)\in \mathbb{R}^d$, $ F({\bm y})\,=\, C\{F_1(y_1),F_2(y_2), \ldots, F_d(y_d) \}.$ Put succinctly, any $d$-dimensional joint distribution can be decomposed into $d$ univariate marginal distributions and a $d$-dimensional copula. 

The application of copulas in finance dates back to \cite{tib96} and \cite{free98} who modelled insurance portfolio risks, and \cite{li00} who priced collateralised debt obligations (CDOs) by modelling the default risks of the assets included in these contracts \footnote{Derivative contracts backed by a pool of loans, the majority of which are mortgages.}.  \cite{emb02} provide an early review the properties and pitfalls of copula modules in investment risk management. The majority of the literature since the 2007–2008 financial crisis\footnote{Often referred to as the great financial crisis.} focused on the inability of Gaussian copulas to capture tail dependence when it came to pricing CDOs and managing investment risk \citep[see][]{salmon09, biel11, zimmer12}. Since then Archimedean copulas were popularised in finance and insurance literature. Archimedean copulas are a family of single parameter copulas characterised by a monotone strictly decreasing convex function referred to as the {\it generator function}. Depending on the choice of {\it generator function} they can capture central dependence only (Frank copula) and either central and lower tail (Clayton copula) or central and upper tail (Gumbel) dependence \citep[see][]{nelsen05, nelsen06, charp09, jaw09, joe14}. Due to the symmetry of Archimedean copulas, margins of the same dimension are equal, a restriction when modelling financial data. \cite{joe97} introduced  hierarchical Archimedean copulas (HACs), obtained by plugging in Archimedean copulas into each other. \cite{mcneil08, mcneil09} argue that HACs can model the asymmetries in the joint dependence structure well and discuss their properties. Examples of their use in finance range from CDO pricing (see \cite{hof11}, \cite{li21}, and \cite{lei23}), insurance risk modelling (see \cite{ alb11}, and \cite{coss18, coss19}), and portfolio management (see, \cite{okh13}, \cite{kakouris14}, \cite{zhu16}, \cite{dewick22}, and \cite{cui24}. 
Vine copulas introduced by \cite{joe94} are an alternative to HACs. Their advantage over HACs is that they can be extended to more than two dimensions through the combination of bivariate parametric copulas with vines (graphical tools used to label covariance constraints). However, due to computational cost , it is very difficult to extend vine copulas to more than five dimensions \citep[see][]{czado19}. Vine copulas have mostly been used in portfolio management see \cite{joe12}, \cite{czado13}, \cite{low13}, and \cite{kraus17}. 
Both HACs and vine copulas are parametric, the degree and direction of tail dependence they can capture depends on the choice of generator function and its parameter(s) as well as the complexity of the data they are applied to, see \cite{simpson21} and \cite{gor24}. These issues coupled with the fact that a copula is a hidden dependence structure make the task of choosing an appropriate parametric copula a non trivial one, see \cite{chen07}. 

Nonparametric copulas, are a more flexible way to model the joint multivariate distribution, as they do not impose strong assumptions on the dependence structure. The main nonparametric methods for copula estimation are kernel density models and polynomial approximations. The issue with kernel density models  is the choice of bandwidth as different levels of smoothing are required near the boundary of the marginals and near the boundary of the copula, \citep[see][]{bow98, muller99, rac15}. For more details on theory, construction, and financial applications of kernel method based copulas \citep[see][]{chen09, ome09, li13, rac15, ma15, gee22}.  Our approach relates to polynomial approximation methods. These date back to \cite{li97}, who showed that copula approximations based on Bernstein polynomials converge faster to the true copula compared to checkerboard and shuffles of min approximations. Orthogonal polynomial expansions (e.g. Legendre, Hermite) can also be used to construct copulas, but they require complex corrections to maintain non-negative density estimates, see \cite{SHIRAYA2024}. Based on \cite{li97} findings \cite{san04} introduced Bernstein copulas and showed that these can achieve a more precise approximation of both known and unknown copulas compared to other nonparametric approximations. For more on properties and extensions of Bernstein copulas  \citep[see][]{san04, san07, bake08, yang15, dou16} and for applications in insurance, financial market behaviour, and financial risk management, \citetext{\citealp{ diers12, boue12, guo17}, \citealp{TAVIN2018}}. 
 
We want to merge the Bernstein polynomial approach to copula estimation to that of infinite mixture models, a popular approach to density estimation in Bayesian nonparametrics. Bayesian nonparametric methods place a prior on an infinite dimensional parameter space and adapt their complexity to the data, see \cite{hjort2010} for a book length review. The literature on building copulas using infinite mixture models is sparse, and most models use Gaussian type copulas (symmetric and skewed) or Gaussian copulas with covariates (conditional copulas) as mixing kernels.
Examples of the former can be found in \cite{wu14} and \cite{wu15}, and of the latter in \cite{dalla18} and \cite{dalla23}. However, none of these models captures tail dependence due to the choice of Gaussian type kernels.
 \cite{shep18} introduced a different Bayesian nonparametric method for modelling copula functions based on Dirichlet Pol\'ya trees. Their approach can extend to more than two dimensions, but the Pol\'ya tree partitions are finite and therefore tail dependence is not captured. 

We propose an alternative Bayesian nonparametric copula model that can capture the asymmetry as well as the central and tail dependence of the joint probability distribution. Our model links the work of \citep{pfe16, pfe17} to that of \cite{burda14}.
\cite{pfe16} and \cite{pfe17} built copula densities driven by discrete distributions using a generalised infinite partition of unity (GPU) approach. GPU copulas can capture asymmetry and tail dependence, depending on the choice of generating function (the probability mass function of discrete distribution). \cite{burda14} use the Bayesian nonparametric approach to introduce a copula method based on a factorisation scheme for multivariate infinite mixture models. They consider univariate Gaussian mixtures for the marginals and a multivariate random Bernstein polynomial copula for the link function, under the Dirichlet process prior. They extend the Bernstein polynomial prior of \citep{petro99a, petro99b} to the multivariate setting, and study the posterior consistency of their proposed model. Although their model extends to more than two dimensions, it does not capture tail dependence due to the Bernstein polynomial prior. This is a hierarchical prior which consists of a random density that is a mixture of beta densities which under certain conditions is the derivative of a Bernstein polynomial. Bernstein polynomials of a fixed order are a family of linearly independent single value polynomials which are a partition of unity on $[0,1]$ \citep[see][]{fitz10}. We utilise these connections between partitions of unity, Bernstein polynomials and Bayesian nonparametric hierarchical priors to build our model. We propose a hierarchical prior which at the lower level of the hierarchy has an approximate GPU copula $c_{\theta}(u,v)=\sum_{i=1}^\infty \sum_{j=1}^\infty \omega_{\theta} \left(i,j\right) \frac{\phi_{i,\theta}(u)}{\alpha_{i,\theta}} \frac{\phi_{j,\theta}(v)}{\alpha_{j,\theta}},$ where  $\phi_{i,\theta}(u)$ is the generating function (chosen to be a discrete distribution) and $\alpha_{i,\theta} = \int_0^1\phi_{i,\theta}(u)du,$ with $\theta$ acting as a smoothing parameter.  
 We place a prior $p(\theta)$ on the smoothing parameter $\theta$ and given $\theta$ the weights $\omega_{\theta} \left(i,j\right)$ are random. We consider two discrete distributions for generating functions, the binomial and the negative binomial. The former choice illustrates the link between partitions of unity and Bernstein polynomials with the resulting copula being the multivariate random Bernstein polynomial copula of \cite{burda14}. Choosing the binomial distribution leads to a copula that can capture asymmetries, but not tail dependence. In contrast, the negative binomial distribution results in a copula that can capture both asymmetries and tail dependence. To proceed to inference we exploit the stick-breaking representation, \citep[see][]{sethuraman94}, of our infinite mixture model and implement a Gibbs sampling scheme based on the slice sampler of \cite{kalli11}. We evaluate the quality of our model on both simulated and real data sets from insurance and finance. The results show that our hierarchical prior with the negative binomial generating function adequately captures the asymmetries and tail dependence of both the insurance and finance data examples, and leads to better out-of-sample predictive performance when compared to popular parametric copula models and to the Bernstein copula of \cite{burda14}.

The paper is organised as follows: Section \ref{S2} reviews the generalised partition-of-unity copula (GPU) model of \citep{pfe16, pfe17} with some examples on the choice of generating functions. Section \ref{S3} presents our Bayesian nonparametric approach based on random GPU copulas under the Dirichlet prior, which we refer to as GPU-Dirichlet copula. We show that the multivariate Bernstein polynomial copula model of \cite{burda14} is a special case of our GPU-Dirichlet copula when the partition is finite and the generating function is the binomial distribution. We explain how infinite partitions of unity allow for asymmetry and tail dependence when the generating function is the negative binomial distribution under the stick-breaking representation of the GPU-Dirichlet prior. In Section \ref{S4} we describe an efficient MCMC algorithm to sample from the posterior based on the stick-breaking representation of the GPU-Dirichlet prior. Section \ref{S5} includes simulation examples to illustrate the ability of our model to capture asymmetries and tail dependencies. In Section \ref{S6} we present empirical illustrations based on two real data sets from insurance and financial econometrics.  Section \ref{S7} concludes with a summary of our model and empirical results followed by a discussion of future work.

\section{Generalised Partitions of Unity (GPU) copulas}\label{S2}

\cite{pfe16, pfe17} introduce a method for constructing bivariate copula functions based on infinite partitions of the unit interval, which they refer to as generalised partition of unity (GPU) copula. GPUs are characterised by a generating function, $\phi_{i,\theta}(u)$, which is the probability mass function (pmf) of a discrete distribution. The choice of pmf affects the copula's ability to capture tail dependence and asymmetries. In real analysis and topology, a partition of unity for a topological space $X$ is a family of real-valued continuous functions $ \mathop{\{f_s\}}\limits_{s \in S}$ which map $X$ to the unit interval $[0,1]$, such that for every point $x\in X$ the $\sum\limits_{s\in S} f_s(x) =1.$ Partitions of unity and Bernstein polynomials (used to construct Bernstein copulas) are linked. Bernstein polynomials of a fixed order $p$ are a family of $p+1$ linearly independent single value polynomials which are a partition of unity for the unit interval $[0,1]$, \citep[see][]{rubin87, fitz10}. This link carries over to copula constructions. In this Section we describe: 1) how to construct GPU copulas, 2) how the choice of generating function, $\phi_{i,\theta}(u)$,  impacts on the GPU's ability to capture tail dependence and asymmetries, 3) show that Bernstein copulas are a sub-class of GPU copulas, and 4) provide a Bayesian perspective to the GPU construction.

We focus on the bivariate case to illustrate the construction of GPU copulas. They can be extended to an arbitrary number of dimensions, and we reserve this for future work. \cite{pfe16,pfe17} define discrete generating functions $\{\phi_i(u)\}_{i \in \mathbb{Z}^+}$ and $\{\psi_j(v)\}_{j \in \mathbb{Z}^+}$ (with parameters $u$ and $v$ respectively) on the unit interval $[0,1]$ for the set of positive integers $\mathbb{Z}^{+} = \{0,1,2,\dots\} $ such that,
 \begin{eqnarray}
 \sum\limits_{i=0}^{\infty} \phi_i(u) = \sum\limits_{j=0}^{\infty}\psi_j(v)=1\label{discrite}\\
\mbox{and}\nonumber \\
\int_{0}^{1}\phi_i(u)du =\alpha_i >0\quad \int_{0}^{1}\psi_j(v)dv =\beta_j >0 \label{priorprob}
 \end{eqnarray}

Eqn (\ref{discrite}) implies that $\phi_i(u)$ and $\psi_j(v)$ can be discrete probability distributions with parameters $u$ and $v$ respectively. Eqn (\ref{priorprob}) leads to the Bayesian interpretation of the sequence of  probabilities $\{\alpha_i\}_{i \in \mathbb{Z}^+}$ and $\{\beta_j\}_{j \in \mathbb{Z}^+}$ as prior predictive distributions since $u\mbox{ and } v$ are defined on $(0,1).$ Prior predictive distributions `exist' on the scale of the data. In fact they are the marginal distributions of the data (normalising constant of the posterior distribution distribution of the parameters) which in this case are $i$ and $j.$ Therefore $\dfrac{\phi_i(u) }{\alpha_i}$ is the posterior distribution of $u$ given $i$ and $\dfrac{\psi_j(v)}{\beta_j}$ is the posterior distribution of $v$ given $j,$ both defined on the unit interval. This links the GPU construction to the beta distribution and as a consequence to Bernstein polynomials and Bernstein copulas, which we describe in Section\ref{S3}.
To complete the construction of the GPU copula \cite{pfe16,pfe17} introduce a sequence of weights  $\{\omega_{ij}\}_{ij\in \mathbb{Z}^+}$ on $\mathbb{Z}^{+}\times \mathbb{Z}^{+}$  which are joint probabilities with marginals, 
$$ 
  \alpha_{i}= \sum_{j=1}^\infty \omega_{ij}, \,\quad\,\beta_{j}= \sum_{j=1}^\infty \omega_{ij} \quad \mbox{where} \quad \sum_{i=1}^\infty\alpha_{i}=\sum_{j=1}^\infty\beta_{j}=1
 $$
Then the bivariate GPU copula with generating functions $\phi_i(u)$ and $\psi_j(v)$ is: 
\begin{equation}
c(u,v) = \sum_{i=1}^{\infty}\sum_{j=1}^{\infty}\omega_{ij} \dfrac{\phi_{i}(u)}{\alpha_{i}}\dfrac{\psi_{j}(v)}{\beta_{j}}\,\, \mbox {for}\,\, u, v \in (0,1)\label{GPU}
\end{equation}
Clearly, Eqn (\ref{GPU}) is a copula density since the marginals $c(u)$ and $c(v)$ are uniformly distributed, i.e.
 $$ c(u)=\int_0^1 c(u,v) dv = \sum_{i=1}^\infty \sum_{j=1}^\infty\frac{\omega_{ij}}{\alpha_i\beta_j}\phi_i(u)\int_0^1 \psi_j(v)dv= \sum_{i=1}^\infty \sum_{j=1}^\infty\frac{\omega_{ij}}{\alpha_i\beta_j}\beta_j\phi_i(u)=\sum_{i=1}^\infty \phi_i(u)=1.$$ In general, Eqn (\ref{GPU}) does not yield a symmetric copula, that is, $c(u,v) \neq c(v,u)$. Symmetry holds if and only if $\phi_i(u) = \psi_i(u)$ and $\omega_{ij} = \omega_{ji}$ for all pairs $(i, j)$. From a Bayesian perspective we can view Eqn (\ref{GPU}), and the GPU copula as the expectation of the posterior distribution over $i\,j$. This is due to the construction of the weights $\omega_{ij}$ and the assumption that parameters $u$ and $v$ are independent and so the $i \,j$ can be generated from the same distribution.

The generating functions $\phi_i(u)$ and $\psi_j(v)$ can be from the same family of discrete probability distributions. We focus on this case and for ease of exposition we will use $\phi_i(u)$.
We follow \cite{mas20} and define the generating function on the set of natural numbers $\mathbb{N}^+ =\{1,2,3,\ldots\},$ shifting by one the usual the support of distributions such as the binomial and the negative binomial. The motivation for choosing $\mathbb{N}^+$ is to clarify the link between GPU and Bernstein copulas. 
The generating functions define the infinite partition on the unit interval $[0,1],$ generated from the set of intervals, $$\Lambda_{i}= \sum_{l=1}^{i}\alpha_{l},\, \, i=1,2,\ldots$$ where $\alpha_l$ is defined by Eqn.(\ref{priorprob}) with $\alpha_0=0$. Therefore, a bivariate partition of the unit square is obtained by, $\left\{(\Lambda_{i-1},\Lambda_{i}]\times (\Lambda_{j-1},\Lambda_{j}]\right\}$, for $i,j=1,2,\ldots$ This infinite partition of the unit square determines the main features of the copula including the presence or absence of tail dependence, which can be measured using the upper and lower tail dependence coefficients, $\lambda_U=\lim\limits_{u \to 1^-} \Pr(V>u|U>u)$ and  $\lambda_L=\lim\limits_{u \to 0^+} \Pr(V\leq u|U\leq u)$, respectively, see \cite{nelsen06}. Choosing $\phi_i(u)$ to be the pmf of a binomial distribution defined on $\mathbb{N}^+$ is what leads to the standard expression of the Bernstein copula, \citep[see][]{san04}. We discuss this in detail in Section 2.1, and consider the GPU copula with $\phi_i(u)$ the pmf of a negative binomial distribution in Section 2.2.

\subsection{Example: Binomial generating function - Bernstein copula}\label{bino} When the generating function $\phi_i(u)$ of the GPU copula is the pmf of the binomial distribution, we have the Bernstein copula of \cite{san04}. The support of the binomial distribution is finite, and so the partition of the unit interval is finite. Let $\phi_i(u)$ be the pmf of a binomial distribution defined on a finite set of positive integers with parameters $\theta, \,\,\mbox{and}\,\, u$ (where $u$ is the probability of 'success' in $\theta$ trials) ,\begin{equation}\label{Bin}
	\phi_{i}(u)= {{\theta-1} \choose{i -1}}u^{i-1}(1-u)^{\theta-i},\quad \text{ for }  i = 1\ldots,\theta.\end{equation} 
The prior predictive distributions (marginals of the data points $i$), $\alpha_i$ , will be equal that is $\alpha_{i}= \int_0^1\phi_{i}(u)du= \frac{1}{\theta}$ for $i = 1\ldots,\theta.$ Then the resulting copula density is, 
	 \begin{equation}\label{BC}
	c(u,v ) = \sum_{i=1}^{\theta} \sum_{j=1}^{\theta}\omega_{ij} \beta(u|i,\theta-i+1)\beta(v|j,\theta-j+1), \end{equation}
 with,\[\sum_{j=1}^{\theta} \omega_{ij} =\sum_{i=1}^{\theta} \omega_{ij} =\frac{1}{\theta},\] since $\alpha_i = \sum_{j=1}^{\theta} \omega_{ij}$ and $\alpha_j = \sum_{i=1}^{\theta} \omega_{ij}.$ 
The copula in Eqn.(\ref{BC}) is a mixture of beta densities $\beta(a,b)$, with scale parameters $(i, \theta-i+1)$ and $(j, \theta-j +1)$ respectively. So choosing the generating function to be the pmf of the binomial distribution, results in Eqn.(\ref{BC})  being is the density of a Bernstein copula, with the weights $\omega_{ij}$ being constant.

\begin{figure}[h!]
\centering
\hspace{0.35in} (a)    \hspace{2in}      (b)\\
\includegraphics[trim= 0mm 15mm 0mm 15mm, clip,scale=0.7]{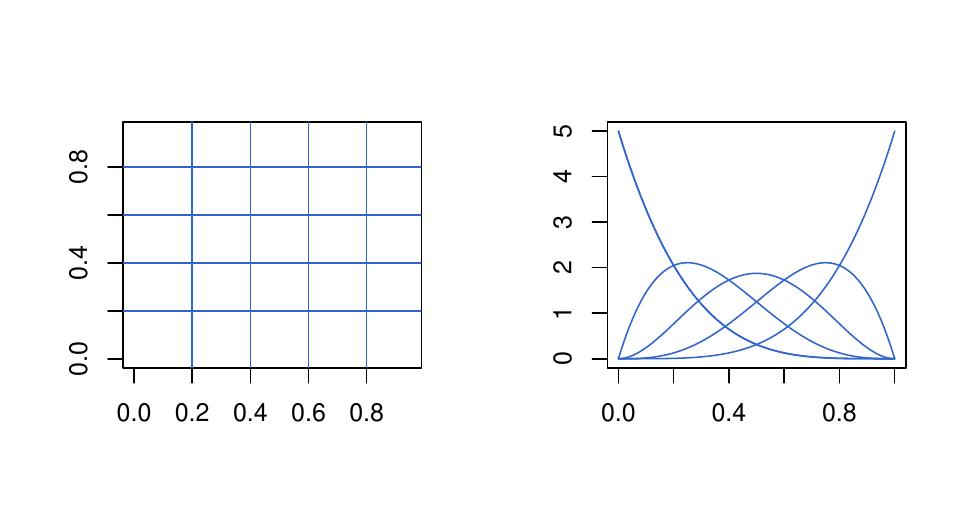} 
\caption{\small  Panel (a) Bernstein partition of the unit square for $\theta= 5$  and panel (b) the corresponding Beta densities, $\beta(u|i,5-i+1)$, for $i=1,\ldots,5$ }\label{BC partition}
\end{figure}

Figure \ref{BC partition} illustrates the finite partition of the bivariate copula Bernstein copula in Eqn (\ref{BC}) with $\theta=5$. We can see that $\theta$ acts as a smoothing parameter of the piecewise constant function defined by the weights, $\omega_{ij}$. The larger the $\theta$ the smoother the copula density. Notice that $|c(u,v)|\leq k$, meaning the Bernstein copula $c(u,v)$ is bounded by $\theta$ which implies that both upper and lower tail dependence are zero.

\subsection{Example: Negative binomial generating function}\label{nbino} 
In this example we let $\phi_i(u)$ be the pmf of a negative binomial distribution with parameters $\theta$ and $u$ (where $u$ is the probability of `success' and $\theta$ is the number of `failures' until a `success') defined on the positive integers as,\begin{equation}\label{NegBin}
	\phi_{i}(u)= \frac{\Gamma(\theta+i-1)}{\Gamma(\theta)\Gamma(i)}(1-u)^\theta u^{i-1},\quad \text { for } i = 1,2,\ldots\end{equation} 
The prior predictive distributions of $i$ are $\alpha_{i}=\int_0^1\phi_{i}(u)du=  \frac{\theta}{(\theta+i-1)(\theta+i)}$, for $i = 1,2,\ldots,$ and the resulting bivariate copula is given by:
\begin{equation}
c(u, v ) = \sum_{i=1}^{\infty} \sum_{j=1}^{\infty}\omega_{ij}\beta(u|i,\theta+1)\beta(v|j,\theta+1),\label{nb}\end{equation}where,  \begin{equation}\label{margs}\sum_{j=1}^\infty \omega_{ij}=\alpha_i=\frac{\theta}{(\theta+i-1)(\theta+i)} \quad \text{and}
	\quad \sum_{i=1}^\infty \omega_{ij}=\alpha_j=\frac{\theta}{(\theta+j-1)(\theta+j)},
\end{equation}
Notice that $c(u,v)$ in Eqn(\ref{nb}) is again a mixture of beta densities with scale parameters $(\theta+i-1, \theta+i)$ and $(\theta+j-1, \theta+j)$ with weights defined in Eqn (\ref{margs}). Parameter $\theta$ controls both the smoothness of the stepwise function $\dfrac{\theta}{(\theta+i-1)(\theta+i)}$, and the level of upper tail dependence $\lambda_{U}(\theta)$. As the value of $\theta$ increases we get smoother weights, and as  $\theta\rightarrow \infty$  we have $ \lim_{\theta\rightarrow\infty}\lambda_{U}(\theta)=1$. A limitation of this GPU copula, pointed out in \cite{pfe16}, is that upper tail coefficients are only known for the diagonal dominance case where all off-diagonal weights are zero, that is  $\omega_{ij}=0$ for $i\neq j$, and $\theta \in \mathbb{N}$. In this case, $$\lambda_U = 1- \frac{{2\theta \choose \theta}}{4^\theta} \approx 1-\frac{1}{\sqrt{\pi\theta}}\text{\quad for large } \theta.$$

\begin{figure}[h!]
\centering
\hspace{0.011in} (a)    \hspace{1.75in}      (b)\\
	\includegraphics[trim= 0mm 15mm 0mm 15mm, clip,scale=0.7]{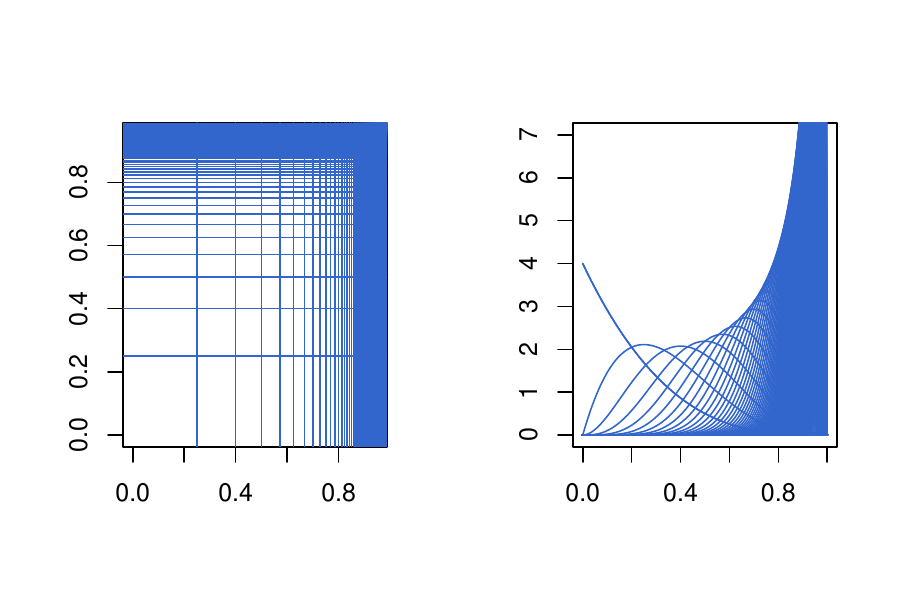} 
		\caption{\small Panel (a) displays the infinite partition for a bivariate GPU copula with a negative binomial generating function and smoothing parameter $\theta=3.$ Panel (b) displays the  corresponding beta densities, $\beta(u|i, 3+1)$, for $i=1,2,\ldots $, allowing for upper tail dependence.}\label{NegativeBN_partition_1}
\end{figure}
\begin{figure}[h!]
\centering 
	\hspace{0.01in} (a)    \hspace{1.75in}      (b)\\
	\includegraphics[trim= 0mm 15mm 0mm 15mm, clip,scale=0.7]{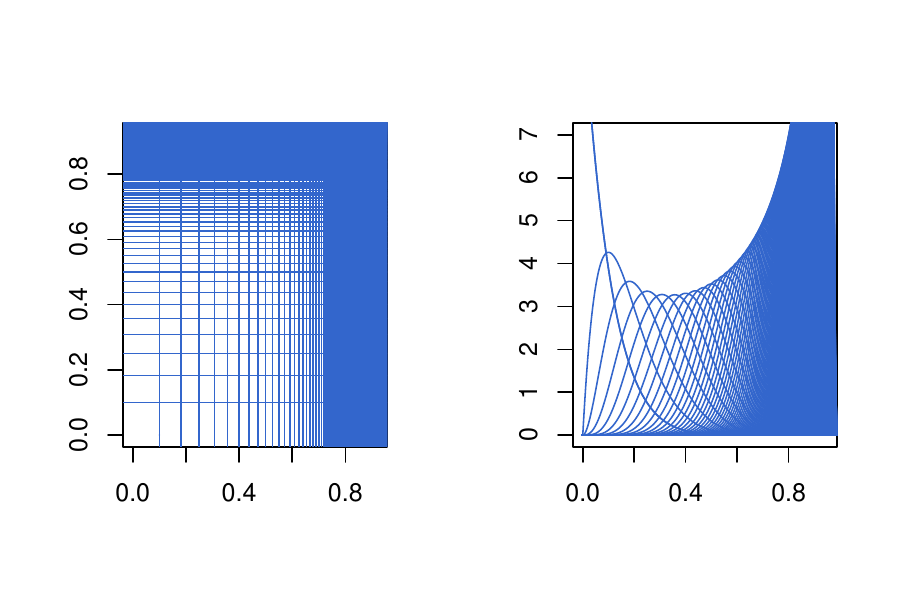} 
	\caption{\small Panel (a) displays the infinite partition for a bivariate GPU copula with a negative binomial generating function and smoothing parameter $\theta=9.$ Panel (b) displays the  corresponding beta densities, $\beta(u|i, 9+1)$, for $i=1,2,\ldots $, allowing for upper tail dependence.}\label{NegativeBN_partition_2}
\end{figure}

Figures \ref{NegativeBN_partition_1} and \ref{NegativeBN_partition_2} display the infinite partition of the bivariate GPU copula with a negative binomial generating function and $\theta = 3$ and $\theta = 9$, respectively. Panels (a) display the infinite partition and panels (b) the corresponding beta densities.  From panels (a) we can see that the step function structure of the weights $\omega_{ij}$, Eqn (\ref{margs}), leads to an infinite partition on the upper right hand side of the unit hyper-square, leading to upper tail dependence, $\lambda_{U}(\theta)$. From panels (b) we can see that the value of $\theta$ affects the magnitude of $\lambda_{U}(\theta)$, as shown by the shape of the corresponding beta densities. 
\bigskip

\vspace{0.25in}

The rotated version of the $c(u,v)$ copula in Eqn(\ref{nb}) leads to lower tail dependence, with density given below:\[
c^{R}(u,v) = \sum_{i=1}^{\infty} \sum_{j=1}^{\infty}w_{ij}\beta(1-u|i,\theta+1)\beta(1-v|j,\theta+1).\]

\section{Random General Partition of Unity Copulas}\label{S3}

In this section, we introduce our model which we refer to as random general partition of unity (RGPU) copula. The weights, $\omega_{ij},$ of the RGPU copula and smoothing parameter $\theta$ of its generating functions are random. We do this by defining a prior based on the Bernstein polynomial prior of \cite{petro99a, petro99b} and its multivariate extension described in \cite{burda14}. We call our novel prior GPU-Dirichlet prior,  and similar to Section \ref{S2} we focus on the bivariate case where the generating functions $\phi_i(\cdot)$ and $\psi_j(\cdot)$ are the same, i.e. $\phi_i=\psi_i.$ We consider two choices for $\phi_i$: \begin{enumerate}\item[a)] $\phi_i(u)$ is the pmf of the binomial distribution which leads to a sub-class of GPU-Dirichlet prior we refer to as BernsteinCBP prior, because it leads to the \cite{burda14} model where the RGPU copula has no tail dependence, and \item[b)] $\phi_i(u) $ is the pmf of the negative binomial distribution which leads to a sub-class of the GPU-Dirichlet prior we refer to as NegBinC prior where the RGPU copula exhibits tail dependence.  \end{enumerate}

\subsection{Constructing a RGPU copula with a GPU-Dirichlet prior}

We define an approximate random bivariate GPU copula with generating function $\phi_{i,\theta}(u)$ as follows,
\begin{equation}c_{\theta}(u,v)
=\sum_{i=1}^\infty \sum_{j=1}^\infty \omega_{\theta} \left(i,j\right) \frac{\phi_{i,\theta}(u)}{\alpha_{i,\theta}} \frac{\phi_{j,\theta}(v)}{\alpha_{j,\theta}},\label{PfeGPU}
\end{equation}
where the smoothing parameter $\theta$ and the mixing weights $\omega_{\theta}\left(i,j\right)$ are random. On each cell of the infinite partition the weights, $\omega_{\theta}\left(i,j\right),$ are given by: 
 \begin{equation}
	\omega_{\theta}\left(i,j\right)=\int_{\Lambda_{i-1}}^{\Lambda_{i}}\int_{\Lambda_{j-1}}^{\Lambda_j}F(u,v)dudv.
\label{randomWGPU}\end{equation} where $F(u,v)$ is a distribution function $F: [0,1]^2\rightarrow 1$ and the breakpoints of the partition are defined as follows, \begin{equation*}
	\Lambda_{i,\theta}= \sum_{l=1}^{i}\alpha_{l,\theta},
	\end{equation*} for any $i\in \mathbb{N}^+ $, where $\alpha_{l,\theta}$ is given by,
	
	\begin{equation*}\label{alpha_j}\alpha_{i,\theta}= \int_0^1\phi_{i,\theta}(u)du.\end{equation*}
	
 
The weights $\omega_{\theta}\left(i,j\right)$ of $c_{\theta}(u,v)$ in Eqn.(\ref{PfeGPU}) are infinite and depend on the choice of smoothing parameter $\theta$. Placing a prior on $\theta$, makes it random which leads to $\omega_{\theta}\left(i,j\right)$ being random too. Consequently the breaks of the partition grid are re-weighted and defined on unequally sized regions. This leads to a flexible data-driven copula model that can capture both central and tail dependence. In the rest of the section we provide the details of this construction.

Recall from Section \ref{S2} that Bernstein polynomials of a fixed order are a partition of unity for the unit interval $[0,1].$  They also have a link to mixture modelling when the kernel of the mixture is the density function of a beta distribution. Our work bridges these results. The Bernstein polynomial (BP) for a bounded mapping $F: [0,1]\rightarrow \mathbb{R}$ with smoothing parameter $\theta$ is given by:
\begin{equation}BP(u|\theta, F)=\sum\limits_{j=0}^{\theta} F\left(\frac{j}{\theta} \right)\binom{\theta}{j}u^j(1-u)^{\theta-j}\,\,\mbox{for} \,\, u\in[0,1]\label{bern}\end{equation}

\cite{mallik94} showed that: 1) when $F$ is a cumulative distribution function then the BP in Eq(\ref{bern}) is also a cumulative distribution function on $[0,1]$ and 2) the derivative of a BP is a mixture of beta densities i.e.  $\sum_{j=1}^{\theta} \omega_{j, \theta}\beta(u;\,a,b)$ where $\beta(a,b)$  is a beta density with parameters $a$ and $b.$ \cite{petro99a, petro99b} showed that when $F(0)=0$ its density function is given by 
\begin{equation} bp(u|\theta, F)=\sum_{j=1}^{\theta} \omega_{j,\theta} \beta(u|j, \theta-j+1),\label{BPP}\end{equation}
where $\omega_{j,\theta} = F(j/\theta)-F((j-1)/\theta).$ When $\theta$ is random with a prior $\pi(\theta)$, the distribution of the weights $\omega_{j,\theta}$ given $\theta$ exits on the simplex $S_{\theta-1}=\{(\omega_1,\ldots, \omega_{\theta}) \in \mathbb{R}^{\theta} \,\,:\,\, 0\leq\omega_j\leq 1, \,\, j=1, \ldots,\theta\,\,\sum_{j=1}^{\theta} \omega_j=1\}$ for every $\theta\in\mathbb{N}^{+}.$ Under this hierarchical set up Eq(\ref{BPP}) is a prior with weak support on the space of all probability measures on $[0,1]$ which \cite{petro99a, petro99b} referred to as the Bernstein-Polynomial prior\footnote{Provided that $\pi(\theta)$ assigns positive mass to all naturals and the density of the weights' distribution is positive.}. She then introduced the Bernstein-Dirichlet (BD) prior with weights  $(\omega_1,\ldots, \omega_{\theta})\sim \text{Dirichlet}(\delta_1,\ldots,\delta_\theta)$, where $\delta_j = M(F_0(j/\theta)-F_0((j-1)/\theta)).$ $F_0$ is a distribution on [0,1] and $M$ a positive constant. The BD prior is equivalent to assuming that $F|M,\,F_0\sim \text{DP}(M,F_0)$ independent of $\theta$, where DP refers to the Dirichlet Process \citep[see][]{ferguson73, ferguson74}. The DP is the most popular Bayesian nonparametric prior due to its properties and the different ways it can be constructed i.e. normalisation of a gamma process, infinite-dimensional generalisation of the Dirichlet distribution, and stick-breaking representation to name a few. In our paper we use the latter two construction methods. For more details on the DP and its representations see \cite{hjort2010}.

We introduce randomness in $c_{\theta}(u,v)$, Eqn. (\ref{PfeGPU}), in two stages: 1) we place a prior, $p(\theta),$ on the smoothing parameter $\theta$, 2) given $\theta$ we assume that the conditional distribution of the mixing weights, $\omega_{\theta}\left(i,j\right)$, is a Dirichlet distribution with parameters $\delta_{i,j}\,\,\mbox{for}\,\, i,j\in \mathbb{N}.$
Following \cite{petro99a, petro99b}, we assume that $\delta_{i,j} = M w_{\theta}^{0} \left(i,j\right)$, where, $$\omega_{\theta}^0\left(i,j\right)=\int_{\Lambda_{i-1}}^{\Lambda_{i}}\int_{\Lambda_{j-1}}^{\Lambda_j}F_0(u,v)dudv.$$ with $F_0$ a distribution function in the unit square and $M $ a positive constant such that $F\mid M,F_0\sim DP(M,F_0)$. We can interpret $M$ as the concentration parameter, controlling how close $F$ is to the base measure $F_0 $. We set $F_0$ to be the uniform distribution and $M= 1$ for both the simulation and real application examples, see Section \ref{S5} and Section \ref{S6}, respectively. Choosing $M=1$ strikes a good balance between having too few or too many mixture components and the uniform base measure avoids strong assumptions about the shape of the distribution.

Similar to the copula construction of \cite{petro99a, petro99b} and \cite{burda14}, given a copula sample, $\left\{(u_1,v_1)\ldots,(u_n,v_n))\right\}$ we introduce a sample of latent variables,  \newline $\left\{(y_{11},y_{1,2})\ldots,(y_{n1},y_{n2}))\right\}$, such that,
 		$$(u_i,v_i)|(y_{i1},y_{i2}) \sim \frac{\phi_{j_1,\theta}(u)}{\alpha_{j_1,\theta}}\times \frac{\phi_{j_2,\theta}(u)}{\alpha_{j_2,\theta}}, \, \text{ if } (y_{i1},y_{i2}) \in (\Lambda_{j_1-1,\theta},\Lambda_{j_1,\theta}]\times(\Lambda_{j_2-1,\theta},\Lambda_{j_2,\theta}]$$
The latent variables, $(y_{i1},y_{i2})$, are interpreted as hidden label variables indicating from which mixture component in Eqn(\ref{PfeGPU}) the variable $(u,v)$ is generated. We can then reformulate the hierarchical prior model as follows,\begin{eqnarray}
	\theta&\sim& p(\theta)\label{hierarchical}\\
	F &\sim& DP(M,F_0)\notag\\
	(y_{i1},y_{i2}) &\sim& F\notag\\
	(u_i, v_i)|(y_{i1},y_{i2}),\theta&\sim&\frac{\phi_{j_1,\theta}(u)}{\alpha_{j_1,\theta}}\times \frac{\phi_{j_2,\theta}(u)}{\alpha_{j_2,\theta}}, \, \text{ if } (y_1,y_2) \in (\Lambda_{j_1-1,\theta},\Lambda_{j_1,\theta}]\times(\Lambda_{j_2-1,\theta},\Lambda_{j_2,\theta}]	\notag
\end{eqnarray} 
Different hierarchical priors are obtained based on the choice of  $\phi_{j,\theta}(u)$. The BernsteinCBP prior of \cite{burda14} is obtained when $\phi_{j,\theta}(u)$ is the binomial pmf of Eqn. (\ref{Bin}). In this case, the partition breakpoints are, $ \Lambda_{j,\theta}=\sum_{l=1}^j \alpha_{l,\theta}=\sum_{l=1}^j \frac{1}{\theta}=\frac{j}{\theta},\, \text{ for }j = 1,\ldots,\theta,$ and the hierachical model becomes, \vspace{-0.2in}

\begin{eqnarray}
		\theta&\sim& p(\theta)	\label{hierarchicalBP}\\
		F &\sim& DP(M,F_0)\notag\\
		(y_{i1},y_{i2}) &\sim& F\notag\\
		(u_i, v_i)|(y_{i1},y_{i2}),\theta&\sim&\beta(u|j_1, \theta-j_1+1)\times \beta(v|j_2, \theta-j_2+1),\notag \\& & \text{ if } (y_{i1},y_{i2}) \in \left(\frac{j_1-1}{\theta}, \frac{j_1}{\theta}\right]\times \left(\frac{j_2-1}{\theta}, \frac{j_2}{\theta}\right]	\notag
	\end{eqnarray} which is the bivariate case of the \cite{burda14} model. Even-though this model allows for asymmetries it does not allow for tail dependence, which is its main drawback.

To obtain the NegBinC prior we choose $\phi_{j,\theta}(u)$ to be the pmf of a negative binomial distribution, Eqn.(\ref{NegBin}). We can derive the expression for the breakpoints, $\Lambda_{j,\theta}$ using the fact that the cumulative distribution function (cdf) of a negative binomial distribution is a regularised incomplete beta function, that is, $$\Lambda_{j,\theta}=\sum_{l=1}^j \alpha_{l,\theta} = \int_0 ^1 \sum_{l=1}^j \frac{\Gamma(\theta+l-1)}{\Gamma(\theta)\Gamma(l)} (1-u)^\theta u^{l-1}du=\int_0 ^1 \frac{\text{B}(1-u; \theta,j)}{\text{B}( \theta,j)}du$$ where $\text{B}(x; a,b)$ is the incomplete beta function,$$\text{B}(x; a,b)=\int_0^x t^{a-1}(1-t)^{b-1}dt,$$ whose indefinite integral is,$$\int\text{B}(x; a,b)dz=x\text{B}(x; a,b)-\text{B}(x; a+1,b),$$ \citep[see][]{pearson68}. We obtain that, $$\Lambda_{j,\theta}=\frac{\text{B}(\theta,j)-\text{B}(\theta+1,j)}{\text{B}(\theta,j)}=1-\frac{\theta}{\theta+j}=\frac{j}{\theta+j}$$
and the hierarchical prior will be,\begin{eqnarray}
	\theta&\sim& p(\theta)	\label{hierarchicalNB}\\
	F &\sim& DP(M,F_0)\notag\\
	(y_{i1},y_{i2}) &\sim& F\notag\\
	(u_i, v_i)|(y_{i1},y_{i2}),\theta&\sim&\beta(u|j_1,\theta+1)\times\beta(v|j_2,\theta+1), \notag\\ && \text{ if } (y_{i1},y_{i2}) \in \left(\frac{j_1-1}{\theta+j_1-1}, \frac{j_1}{\theta+j_1}\right]\times\left(\frac{j_2-1}{\theta+j_2-1}, \frac{j_2}{\theta+j_2}\right]	\notag
\end{eqnarray} leading to a Bayesian nonparametric copula with both asymmetry and  upper tail dependence. Clearly, the rotated version of this model allows for lower tail dependence.

\subsection{Stick-breaking representation}

Our general hierarchical model in Eqn(\ref{hierarchical}) can be expressed as an infinite mixture since $F$ follows a DP. This infinite mixture is similar to the Dirichlet process mixture (DPM) model defined by \cite{ferguson83} and \cite{lo84}, where the latent variables $(y_{1},y_{2})$ are the conditioning parameters from a copula distributed according to a DP. We can therefore rewrite our hierarchical model in Eqn(\ref{hierarchical}) using the stick-breaking representation \citep[see][]{sethuraman94} as follows,  
\begin{equation}\label{randomGPU_stick}
c(u,v\mid \theta,\boldsymbol{\rho},\mathbf{y})=\sum_{s=1}^{\infty}\rho_{s} \frac{\phi_{h_\theta(y_{s1}),\theta}(u)}{\alpha_{h_\theta(y_{s1}),\theta}}\times \frac{\phi_{h_\theta(y_{s2}),\theta}(v)}{\alpha_{h_\theta(y_{s2}),\theta}}, \end{equation} with $$h_\theta(y)=j, \quad  \text{if }\quad y\in\left( \Lambda_{j-1,\theta},\Lambda_{j,\theta}\right], \quad \text{for }\quad  j \in \mathbb{N},$$  
where $\theta\sim p(\theta)$, and $\boldsymbol{\rho}=\left\{ \rho_s\right\}_{s=0}^\infty$, such that $\rho_1 = \nu_1$ and  $\rho_s =\nu_s\prod_{l=1}^{s-1}(1-\nu_l)$,  with $\nu_s\sim\beta(1,M)$, for  $s=2,\ldots$, and $\mathbf{y}=\left\{(y_{s1},y_{s2})\right\}_{s=1}^\infty$, with $(y_{s1},y_{s2})\sim F_0$. 
\smallskip

This means that the hierarchical BernsteinCBP prior in Eqn(\ref{hierarchicalBP}), proposed in \cite{burda14}, can be  expressed as the following  infinite mixture, \begin{equation}\label{B-stick}
c(u,v\mid \theta,\boldsymbol{\rho},\mathbf{y})=\sum_{s=1}^{\infty}\rho_{s}\beta\left(
u|h_\theta\left( y_{s1}\right),\theta-h_\theta\left( y_{s1}\right)+1\right) 
\beta\left(v|h_\theta\left( y_{s1}\right),\theta-h_\theta\left( y_{s2}\right)+1\right)  \end{equation} where,  $$h_\theta(y)=j, \quad \text{if } y\in\left(\frac{j-1}{\theta},\frac{j}{\theta}\right], \quad \text{for } i= 1,2,\ldots,\theta.$$  
The same is true for the NegBinC prior (\ref{hierarchicalBP}), which leads to our nonparametric copula proposal allowing for tail dependence. Its stick-breaking representation is as follows, \begin{equation}\label{NB-stick}
c(u,v\mid \theta,\boldsymbol{\rho},\mathbf{y})=\sum_{s=1}^{\infty}\rho_{s}\beta\left(
u|h_\theta\left( y_{s1}\right),\theta+1\right)\beta\left(
v|h_\theta\left( y_{s2}\right),\theta+1\right)   \end{equation}  
where, $$h_\theta(y)=i, \quad \text{if } y\in\left(\frac{i-1}{\theta+i-1},\frac{i}{\theta+i}\right], \quad \text{for } i= 1,2,\ldots$$

\section{Computational implementation}\label{S4}

In this section, we describe how (given a copula data sample, $  \left\{(u_{i},v_{i})\right\}$, for $i = 1,\ldots,n$) we sample from the posterior of our model in Eqn(\ref{randomGPU_stick}) defined in Section \ref{S3}. 
For clarity of exposition we consider the case with the NegBinC prior, Eqn(\ref{NB-stick}), however an analogous procedure can be used for the BernsteinCBP prior, Eqn(\ref{B-stick}). Our Gibbs sampler is based on the slice sampler of \cite{kalli11}, an MCMC algorithm used for fitting infinite mixture models with a wide range of stick-breaking priors.  The sampler relies on the automatic truncation of the infinite mixture through the use of latent uniform random variables, which determine the minimum number of mixture components required to proceed with the chain.  We introduce the uniform variables  
$v_i$, for $i= 1,\ldots,n,$  which transform the infinite mixture in Eqn (\ref{randomGPU_stick}) into a finite mixture representation as follows,\begin{eqnarray*}
	p(u_i, v_i,\nu_i\mid \boldsymbol{\theta},\boldsymbol{\rho},\mathbf{y})&=&\sum_{s=1}^{\infty}\mathbf{1}(\nu_i<\rho_s)
	\beta\left(u_i|h_\theta\left( y_{s1}\right),\theta+1\right)
	\beta\left(
	v_i|h_\theta\left( y_{s2}\right),\theta+1\right)\\
	&=&\sum_{s\in A_{\boldsymbol{\rho}}(v_i)}	\beta\left(u_i|h_\theta\left( y_{s1}\right),\theta+1\right)
	\beta\left(
	v_i|h_\theta\left( y_{s2}\right),\theta+1\right),
\end{eqnarray*}
where the set $ A_{\boldsymbol{\rho}}(v_i)=\left\{ s: v_i<\rho_s\right\}$ is finite. To indicate which mixture component the observation $\mathbf{u}_i$ belongs to we 
 introduce set of  auxiliary allocation variables, $z_i$,  for $i= 1,\ldots,n,$ as follows,\[
p(u_i, v_i, \nu_i, z_i\mid \boldsymbol{\theta},\boldsymbol{\rho},\mathbf{y})=\mathbf{1}(v_i<\rho_{z_i})
\beta\left(u_i|h_\theta\left( y_{z_i1}\right),\theta+1\right)
\beta\left(
v_i|h_\theta\left( y_{z_i2}\right),\theta+1\right).
\] This leads to the complete likelihood for the augmented data sample given by,\[l( \boldsymbol{\theta},\boldsymbol{\rho},\mathbf{y}\mid
\mathbf{u}, \mathbf{v}, \mathbf{z})=\prod_{i=1}^n\mathbf{1}(v_i<\rho_{z_i})
\beta\left(u|h_\theta\left( y_{z_i1}\right),\theta+1\right)
\beta\left(
v|h_\theta\left( y_{z_i2}\right),\theta+1\right)
\]where $\mathbf{u_i}=\left\{(u_1,v_1),\ldots,(u_n,v_n)\right\}$, $\mathbf{v_i}=\left\{v_1,\ldots,v_n\right\}$ and $\mathbf{z}=\left\{z_1,\ldots,z_n\right\}$. Following \cite{kalli11} the steps of our MCMC are:
\begin{enumerate}
		\item Sample a finite number of weights $(\rho_1,\ldots,\rho_{s^{\ast}})$ jointly with $(v_1,\ldots,v_n)$ using:\begin{enumerate}\item Sample from $\eta_s\sim \beta(n_s+1,n-\sum_{l=1}^s n_l+M)$ for $s=1,\ldots,z^{\ast}$, where $z^{\ast}=\max\{z_1,\ldots,z_n\}$ and $n_s = \sum_{i=1}^n \mathbf{1}(z_i=s)$ and set $\rho_s  = \eta_s \prod_{l=1}^{s-1}(1-\eta_{l})$.
			\item Sample $v_i$ by simulating from $U(1,\rho_{z_i})$ for $i=1,\ldots,n$.
			\item If necessary, generate more weights, $\rho_s$, from the prior, by simulating from $\eta_s\sim \beta(1,M)$, until $\sum_{s=1}^{s^{\ast}} \rho_s>1-v^{\ast}$, where $v^{\ast}=\min{\{v_1,\ldots,v_n\}}$.
		\end{enumerate}
		\item For $s=1,\ldots,s^{\ast}$, sample independently the mixture parameters, $y_{s1}$ and $y_{s2}$,  by simulating from:\begin{equation}\label{post y_{s1}}
			p(y_{s1}|\cdots)\varpropto \prod_{i:z_i=s} \beta\left(u_i|h_\theta\left( y_{s1}\right),\theta+1\right), \quad 0<y_{s1}<1.
		\end{equation}\begin{equation}\label{post y_{s2}}
		p(y_{s2}|\cdots)\varpropto \prod_{i:z_i=s} \beta\left(v_i|h_\theta\left( y_{s2}\right),\theta+1\right), \quad 0<y_{s2}<1.
		\end{equation}If none of the $z_i$, for $i=1,\ldots,n,$ are equal to a particular mixture component, $s$, then  $\mathbf{y}_s$ is sampled from $F_0$.
		\item For $i=1,\ldots,n$, sample the allocation variables, $z_i$,  by simulating from:
		\begin{equation}\label{z_i}
			P(z_i=s|\cdots)\propto  \mathbf{1}(v_i<\rho_s) 
			\beta\left(u_i|h_\theta\left( y_{s1}\right),\theta+1\right)
			\beta\left(v_i|h_\theta\left( y_{s2}\right),\theta+1\right), 
		\end{equation}for  $s \in \left\{s: v_i<\rho_s\right\}.$
		\item Sample $\theta$ from the following conditional distribution:
		\begin{align}\label{post theta}
			p(\theta|\cdots)\propto ~& p(\theta)\prod_{i=1}^n 
			\beta\left(u_i|h_\theta\left( y_{s1}\right),\theta+1\right)
		\beta\left(v_i|h_\theta\left( y_{s2}\right),\theta+1\right).
		\end{align}
\end{enumerate}

Sampling the label variables, $z_i$, in Eqn(\ref{z_i}) is straightforward since given $v_i$, only a finite number of mixture components are possible. The posterior distributions for $y_{s1}$ and $y_{s2}$ shown in Eqn(\ref{post y_{s1}}) and Eqn(\ref{post y_{s2}}) respectively,  are stepwise constant density functions on $[0,1]$ with an infinite number of intervals. Although sampling algorithms for discrete distributions with infinite support can be used to sample $y_{s1}$ and $y_{s2}$, we decided to use the adaptive random walk Metropolis–Hastings algorithm of \cite{atcha05}. We found that the adaptive random walk Metropolis Hastings algorithm performs better. It has a shorter run time compared to sampling from a discrete distribution with infinite values. The latter can be very time consuming if the mass probabilities are very small. For more details on adaptive MCMC \citep[see][]{andreu08}. We also used the adaptive random walk Metropolis–Hastings algorithm to sample from the posterior distribution of $\theta$ given in Eqn(\ref{post theta}). Recall that we choose the base measure  $F_0$ to be the uniform distribution, but different options could be considered. We note that our MCMC scheme can be scaled to more than two dimensions. This is because copula observations are assumed to be independent given the mixture component they belong to which means that model parameters can be sampled in parallel. In this paper we consider bivariate copulas and assume the same smoothing parameter, $\theta$, for all margins which can be restrictive because it implies that the tail dependence is the same for all pairs of variables. We extend our model to higher dimensions and consider different smoothing parameters for the margins in our future work. 

\section{Simulation study}\label{S5}
In this section we discuss the results of two simulation experiments. The first experiment focuses on capturing tail dependence, by generating data from various parametric Archimedean copulas that exhibit upper tail dependence, lower tail dependence and no tail dependence (under two different levels of Kendall's tau correlation). The second experiment focuses on capturing both tail dependence and asymmetries by generating data from a mixture of two distributions with standard Gaussian marginals and two different bivariate Archimedean copulas. To demonstrate the superior performance of our proposed NegBinC copula over the BernsteinCBP copula of \cite{burda14} in capturing tail dependence and asymmetries we, 1) present scatter plots of predictive samples and compare them with the corresponding simulated data, and 2) compute log-predictive scores (LPS) using simulated out-of-sample data of the same sizes as the in-sample data. Given the training data, $\left\{u_i,v_i\right\}_{i=1}^n$, the LPS for the test data is calculated as follows,  $\left\{u_i,v_i\right\}_{j=n+1}^{2n}$, as follows:\[
\text{LPS}= \sum_{j = n+1}^{2n}\log c\left(u_j,v_j \mid \left\{u_i,v_i\right\}_{i=1}^n\right) \]where, for each $j = n,\ldots,2n,$ \begin{eqnarray*}	
c\left(u_j,v_j \mid \left\{u_i,v_i\right\}_{i=1}^n\right)&\approx& \frac{1}{T}\sum_{t=1}^T c\left(u_j,v_j \mid\theta^{(t)},\boldsymbol{\rho}^{(t)},\mathbf{y}^{(t)}\right) \\
&=& \frac{1}{T}\sum_{t=1}^T \sum_{t=1}^{s^{*(t)}} \rho_{s}^{(t)} \frac{\phi_{h_{\theta^{(t)}}\left(y_{s1}^{(t)}\right),\theta^{(t)}}(u)}{\alpha_{h_\theta^{(t)}\left(y^{(t)}_{s1}\right),\theta^{(t)}}}\times \frac{\phi_{h_{\theta^{(t)}}\left(y_{s2}^{(t)}\right),\theta^{(t)}}(v)}{\alpha_{h_\theta^{(t)}\left(y^{(t)}_{s2}\right),\theta^{(t)}}} \end{eqnarray*}
with $\left\{\theta^{(t)},\boldsymbol{\rho}^{(t)},\mathbf{y}^{(t)}\right\}$ being a posterior draw of the MCMC posterior sample of size $T$. Clearly, models with larger LPS are preferred. 

We also use 1) scatter plots of predictive samples and 2) LPS  to assess the performance of our model when we apply it to real data sets, see Section \ref{S6}.

\subsection{Experiment I - Capturing tail dependence}
\begin{figure}
	\begin{center}\includegraphics[scale=0.9]{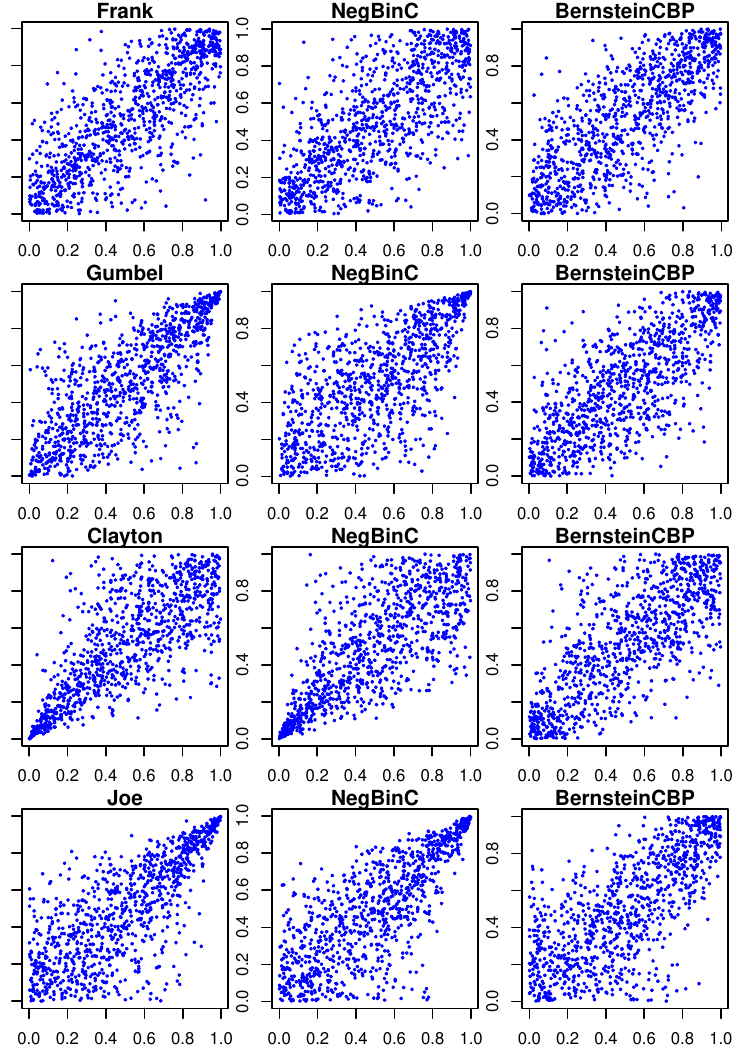}\end{center} 
	\caption{\small Simulated data from parametric copulas (left column) with $n=1000 $ and $\tau =0.6$ and predictive samples of the same size obtained with our proposed NegBinC model (middle column) and the BernsteinCBP model by \cite{burda14} (right column).}\label{Sim1}
\end{figure}

For this experiment we simulate data from the Frank, Gumbel, Clayton and Joe copulas with sample sizes equal to $n =500$, $1000 $ and $3000$.  To assess the ability of our model to capture different levels of tail dependence we consider Kendall's $\tau$ rank correlations equal to $0.3$ and $0.6$. We fit our random GPU copula model under both cases of the GPU-Dirichlet prior: the BernsteinCBP prior by \cite{burda14} and our proposed NegBinC prior which allows for tail dependence.  We apply the rotated version of the NegBinC prior when required.We set $F_0$, the base measure, to be the uniform distribution and the precision parameter  $M= 1.$ Choosing $M=1$ strikes a good balance between having too few or too many mixture components and the uniform base measure avoids strong assumptions about the shape of the distribution. We run our proposed MCMC algorithm for 20000 iterations discarding the first 10000 iterations as burn-in.
 Table \ref{LPSSim1} displays the LPS values obtained for the NegBinC and the BernsteinCBP models under two levels of Kendall's $\tau,$ $0.3$ and $0.6$, for the three different sample sizes $n=500,$ $1000,$ and $3000$ for the data simulated from the Frank, Gumbel, Clayton and Joe copulas. The values in bold indicate the preferred model. Clearly the NegBinC model outperforms the BernsteinCBP  model for both levels of Kendall's $\tau$ and all sample sizes for data simulated from the three copulas with tail dependence (Gumbel, Clayton and Joe). These results verify what we have observed in Figure \ref{Sim1}. For the Frank copula, which has zero tail dependence, the BernsteinCBP  only marginally outperforms the NegBinC model at sizes  $n=500$ and $3000$ for $\tau=0.3$ and sizes  $n=1000$ and $3000$
for $\tau=0.6.$
\begin{table}
	\begin{center}
		\begin{tabular}{ |c|c|l|c|c|c|c| } 
		
				\hline
			& &  & Frank  & Gumbel & Clayton  & Joe \\
			\hline
			\multirow{2}{4em}{$\tau = 0.3$}&
			\multirow{2}{4em}{$n=500$}& NegBinC&0.0743 & {\bf 0.1289} &{\bf 0.1422 } &{\bf 0.1649} \\ 
			& & BernsteinCBP& {\bf 0.0861} & 0.1191 & 0.1362 &0.1389\\ 
			\hline
			&\multirow{2}{4em}{$n=1000$}  &  NegBinC &{\bf 0.0937} & {\bf 0.1400} & {\bf 0.1436} &{\bf 0.1484} \\ 
			& & BernsteinCBP &0.0844 & 0.1327 & 0.1292 & 0.1347\\  
			\hline
				
			&\multirow{2}{4em}{$n=3000$}& NegBinC & 0.0997& {\bf 0.1260} &{\bf 0.1411} &{\bf 0.1575}\\ 
			&& BernsteinCBP & {\bf 0.1067}&0.1168  & 0.1327 &0.1365 \\ 
			\hline
			\hline

			\multirow{2}{4em}{$\tau = 0.6$}&
			\multirow{2}{4em}{$n=500$}& NegBinC &{\bf 0.3891}  & {\bf 0.5682 } & {\bf 0.5947} &{\bf 0.6158}\\ 
			& & BernsteinCBP& 0.3835 &0.4248  &0.4289 &  0.5132 \\ 
			\hline
			&\multirow{2}{4em}{$n=1000$}  & NegBinC & 0.4634&  {\bf 0.5458}&  {\bf 0.6237}& {\bf 0.5773} \\ 
			& & BernsteinCBP &{\bf 0.4804} & 0.4603 &0.4273  & 0.4176 \\  
			\hline
			
			&\multirow{2}{4em}{$n=3000$}& NegBinC &0.4664 &{\bf 0.5306} &  {\bf 0.6130}&{\bf 0.5821}\\ 
			&& BernsteinCBP &{\bf 0.4924} &0.4435  & 0.4455 &0.4885 \\ 
			\hline

		\end{tabular}
	\end{center}
	\caption{LPS values obtained for our proposed NegBinC model and the BernsteinCBP model by \cite{burda14}  under two levels of Kendall's tau $0.3$ and $0.6$, for the three different sample sizes $n-500,$ $1000,$ and $3000$ for the data simulated from the Frank, Gumbel, Clayton and Joe copulas.}
	\label{LPSSim1}
\end{table}

Figure \ref{Sim1} compares the simulated copula data with sample size $n = 1000$ for the four Archimedean copula models with  $\tau = 0.6$ (left column) with predictive samples of the same size obtained from the MCMC algorithm defined in Section \ref{S4} for the  NegBinC model (middle column) and BernsteinCBP model (right column). We can see that the BernsteinCBP model is not capturing the tail dependence of the Clayton, Gumbel and Joe copula simulated data. However, it is clear that our NegBinC  model adequately captures the copula data distribution and, in particular, the tail behaviour for all four different Archimedean copulas.  We omit the results for sample sizes $n = 500$ and $n = 3000$ as well as for  $\tau = 0.3 $ under sample sizes $n =500$, $1000 $ and $3000$ because they are very similar (they can be available upon request). 

\subsection{Experiment II - Capturing asymmetry and tail dependence}

For our second experiment we simulate skewed copula data with tail dependence to illustrate the ability of our random GPU copula model with the NegBinC prior in capturing both asymmetry and tail dependence. The dependence structure of our simulated skewed copula data appears frequently in mixtures of distributions but cannot be captured by mixtures of parametric copulas. Bernstein copulas like the BernsteinCBP of \cite{burda14} can approximate dependence asymmetries but not in presence of tail dependence. 

We simulate data from a mixture of two distributions with standard Gaussian marginals and two different bivariate copulas, a Clayton and a Gaussian copula, as follows: \begin{equation}\small \label{mixt}F(x_1,x_2)=\omega C_\gamma(\Phi(x_1|\mu_{11},\sigma_{11}),\Phi(x_2|\mu_{12},\sigma_{12})) + (1-\omega) C_\rho(\Phi(x_1|\mu_{21},\sigma_{21}),\Phi(x_2|\mu_{22},\sigma_{22})),
\end{equation}
where $C_\gamma$ and $C_\rho$ denote the Clayton and Gaussian copulas with parameters $\gamma$ and $\rho$, respectively;  $\Phi(x|\mu,\sigma)$ denotes a Gaussian distribution with mean $\mu$ and standard deviation $\sigma$; and $\omega$ denotes a weight in $(0,1)$. The second mixture component with Gaussian copula and Gaussian marginals corresponds to a standard bivariate Gaussian distribution with correlation, $\rho$. We obtain skewed bivariate copula data by simply applying the marginal mixture cumulative distribution functions,\begin{eqnarray}\label{copmixt}u_1 = &\omega \Phi(x_1|\mu_{11},\sigma_{11})+(1-\omega) \Phi(x_1|\mu_{21},\sigma_{21}) \\u_2 = &\omega \Phi(x_2|\mu_{12},\sigma_{12})+(1-\omega) \Phi(x_2|\mu_{22},\sigma_{22})
\end{eqnarray} 

\begin{figure}
	\begin{center}\includegraphics[scale=0.70]{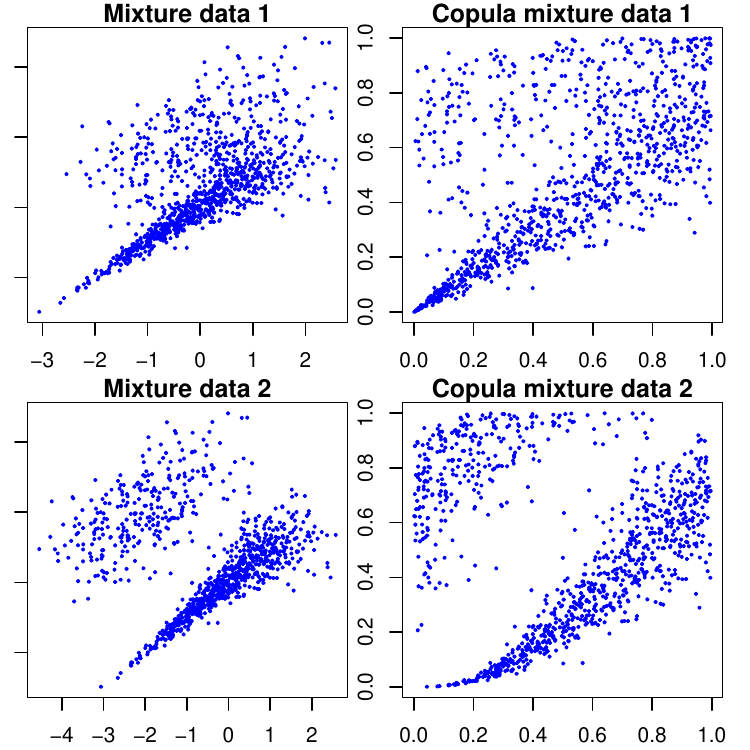}\end{center} 
		\caption{\small Simulated data from (\ref{mixt}) (left column) with $\gamma= 6$; $\rho = 0.6$; $(\mu_{11},\mu_{12})=(0,0)$ and two different means $(\mu_{21},\mu_{22})=(0,2)$ and $(\mu_{21},\mu_{22})=(-2,2)$ (first and second row, respectively) and corresponding copula data using Eqn.(\ref{copmixt}) and Eqn.(23) (right column) using $n=1000$.}\label{Sim2}
\end{figure}

Note that with the representation of $(u_1,u_2)$ in Eqn(\ref{copmixt}) and Eqn(23) we have a copula variable that is skewed and shows tail dependence but it does not follow a mixture of a Clayton and a Gaussian copula. To illustrate our point, we simulate data of sample size $n = 1000$ from this copula model assuming parameters $\gamma= 6$; $\rho = 0.6$; $(\mu_{11},\mu_{12})=(0,0)$ and two different means $(\mu_{21},\mu_{22})=(0,2)$ and $(\mu_{21},\mu_{22})=(-2,2)$.  All standard deviations are set equal to one. Figure \ref{Sim2} displays the simulated data from Eqn(\ref{mixt}) for the two sets of parameters in the left column and the corresponding copula data in the right column. Notice that the copula data show both skewness and lower tail dependence. They are also asymmetric with respect to the diagonal axis, something that does not occur with a mixture of a Clayton and Gaussian copulas (we have symmetry with respect to the diagonal axis). We fit the two considered GPU-Dirichlet models using our proposed MCMC sampler for 20,000 iterations discarding the first 10,000 as burn-in.

\begin{figure}
	\begin{center}\includegraphics[scale=0.70]{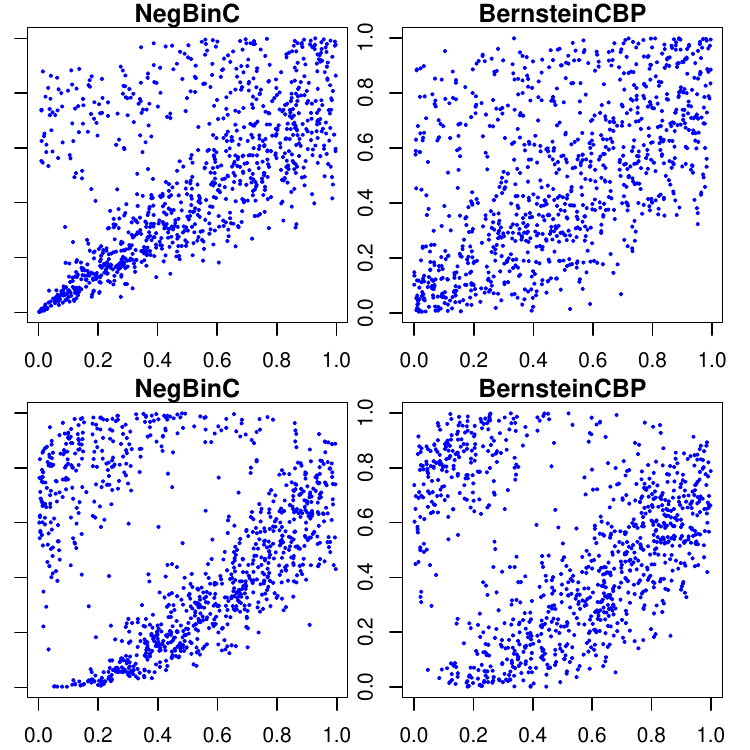}\end{center} 
	\caption{\small Predictive samples of the same size as the simulated data in Figure \ref{Sim2} obtained with the 
		proposed GPU-Dirichlet model using a negative binomial generating function,  the NegBinC model 
		(left) and a binomial generating function, the BernsteinCBP model (right) by \cite{burda14}.}\label{Sim2bis}
\end{figure}

Figure \ref{Sim2bis} displays the comparison of the predictive distributions of the NegBinC model (left column) and BernsteinCBP model (right column). Both models capture the skewness in the copula distribution, however the NegBinC model clearly outperforms the BernsteinCBP model because it captures the strong dependence in the left tail. Finally, Table \ref{LPSSim2} compares the LPS values based on a set of 1000 out-of-sample simulated data for both models. The proposed NegBinC  model clearly outperforms the BernsteinCBP  model by \cite{burda14}.

\begin{table}
	\begin{center}
		\begin{tabular}{ |l|c|c| } 
			\hline
		     &  Copula mixture 1  & Copula mixture 2 \\
			\hline
			 NegBinC & {\bf 0.6081} & {\bf 0.6693} \\ 
			 BernsteinCBP & 0.3874 & 0.5248 \\  
			\hline	
		\end{tabular}
	\end{center}
	\caption{\small LPS values obtained for the simulated parametric copulas (columns) with the 
		proposed GPU-Dirichlet model using a negative binomial generating function,  the NegBinC model 
		(left) and a binomial generating function, the BernsteinCBP model (right) by \cite{burda14} with $n = 1000$.}
	\label{LPSSim2}
\end{table}

\section{Real applications}\label{S6}

\subsection{Insurance data}

In insurance risk management, insurers want to mitigate their losses, it is therefore important to have a model that adequately captures tail dependence of claims variables. The insurance data set we analyse in this section was obtained from the Mendeley data repository \citep[see][]{claims}. It comprises of $n = 1000$ individual insurance policies where claims were made, collected from various insurance companies across three U.S. states: South Carolina, Virginia, and New York. We are interested in exploring the relationships between vehicle, property, and injury claims. The data set covers a period from 1990 to 2015, and we use the data recorded prior to 2008 as in-sample observations for estimation, and reserve the data from 2008 onwards for out-of-sample predictive performance evaluation.

\begin{figure}[h!]
\centering
\includegraphics[scale=0.8]{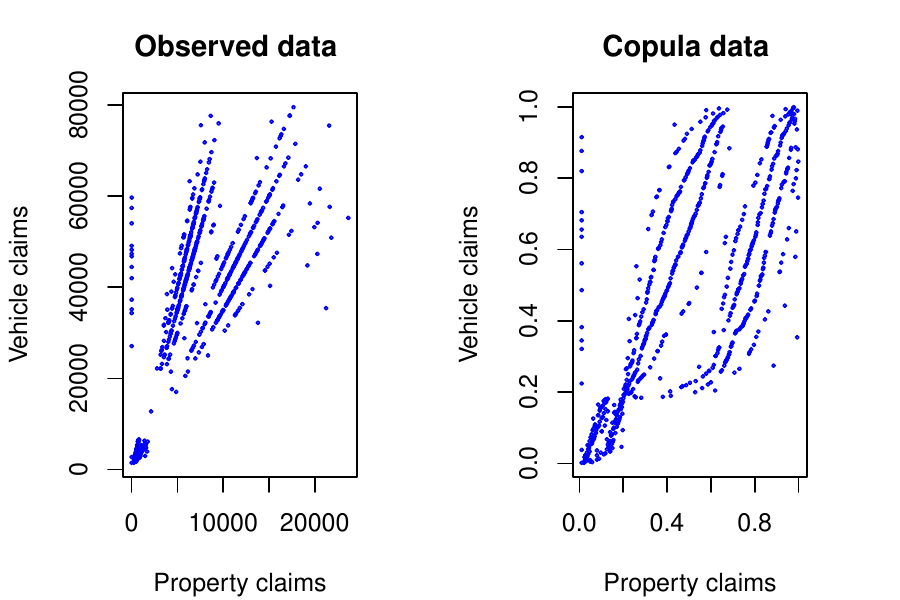}
	\caption{\small Scatterplots of the relationship between the  property and vehicle claims (left) and pseudo-observations with uniform margins (right). }\label{Claims}
\end{figure}

Figure \ref{Claims} displays the scatter plot of the relationship between observed (in-sample) property and vehicle claims (left) and the scatter plot of the copula data using ``pseudo observations'' (right). This illustrates the strong tail dependence between these two types of claims. The observed data plot reveals a positive, heterogeneous and heteroskedastic relationship, characterised by lower tail dependence. The sample Kendall's $\tau$ rank correlation coefficient is $0.5514$. Based on these findings we generate the copula data using ``pseudo-observations'', which are normalised ranked data with uniformly distributed margins displayed in Figure \ref{Claims} (right). It is clear that a similar strong level of asymmetry in the dependence is evident.

\begin{figure}[h!]
\centering
\includegraphics[scale=0.7]{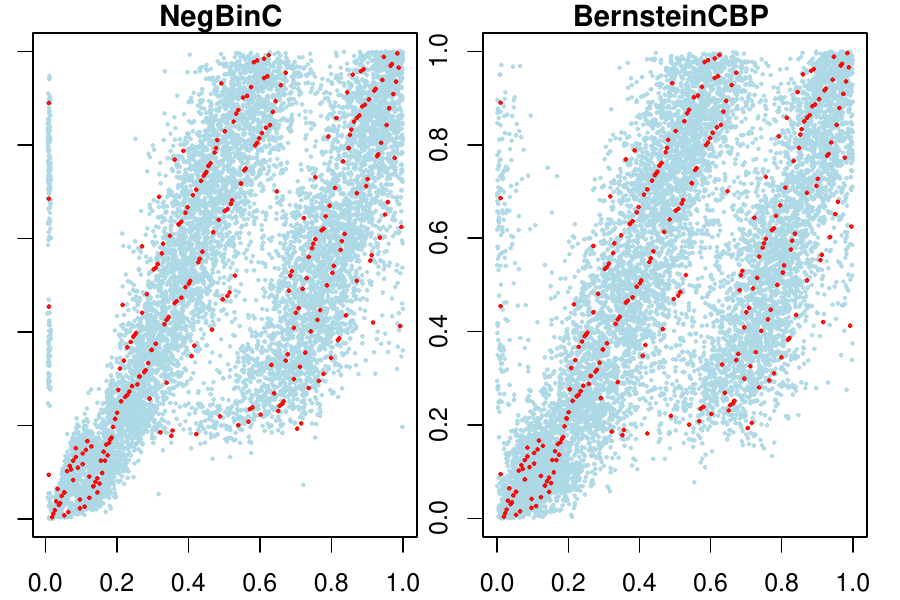}
	\caption{\small Predictive samples (blue) obtained from the proposed GPU-Dirichlet model using a negative binomial generating function,  the NegBinC model (left) and a binomial generating function, the BernsteinCBP model (right) of \cite{burda14}, together with the out-of-sample pseudo-observations (red) for the property and vehicle claims.}\label{Claims2}
\end{figure}

Using our MCMC sampler (with the same specifications as those of the simulated examples), we fit the proposed GPU-Dirichlet model using 1) a negative binomial generating function ie the NegBinC model and 2) a binomial generating function ie the BernsteinCBP model of \cite{burda14} to each pair of variables. Figure \ref{Claims2} displays the predictive samples generated by the NegBinC model (left) and the BernsteinCBP model (right) with the out-of-sample pseudo-observations for property and vehicle claims. The predictive samples are in blue and the out-of-sample pseudo-observations in red. Both Bayesian nonparametric models capture the skewness in the dependence. However, the NegBinC model is tighter at the lower tail demonstrating its ability to adequately capture lower tail dependence and the presence of zeros in the left tail.

Table \ref{LPSClaims} displays the LPS values obtained for the out-of-sample data for each pair of insurance claims, vehicle and property, vehicle and injury, and injury and property. We compare six models, the two Bayesian nonparametric models, NegBinC and BernsteinCBP, three Archimedean copulas, rotated Gumbel, Clayton and rotated Joe, and the Gaussian copula. The values in bold highlight the best model for each pair. The two Bayesian nonparametric models outperform the parametric models, as they allow for heterogeneous and asymmetric dependence. The best model is the NegBinC due to its ability to capture lower tail dependence. This suggests that the  NegBinC model provides a more accurate representation of the central distribution, asymmetries and tail dependence.

\begin{table}
	\begin{center}
		\begin{tabular}{|l |c |c|c|c|c|c|c| } 
			\hline
		&	NegBinC &  BernsteinCBP  & RotGumbel & Clayton  & RotJoe & Gaussian \\
			\hline
		Vehicle \& Property &{\bf	0.7481 }& 0.7164   & 0.3684 & 0.3484& 0.3496& 0.2904\\
			\hline 		
			
		Vehicle \& Injury  &{\bf 0.7973} & 0.7486   & 0.4046 &0.3561 &0.3611 & 0.3588\\
	\hline 	
	Injury \& Property  &{\bf 0.9953} &  0.5882  &0.2821  & 0.2625 &0.2737 &0.2315\\
	\hline 				
		\end{tabular}
	\end{center}
	\caption{ \small LPS values for pairs of insurance claims for the two Bayesian nonparametric copula models, the three Archimedean copulas and the Gaussian copula.}
	\label{LPSClaims}
\end{table}

\subsection{Financial time series data}

The stock returns of publicly traded companies within the finance sector are highly correlated. For our second empirical illustration we  apply our model to the daily returns of three commercial and investment bank stocks: Hong-Kong and Shanghai Banking Corporation (HSBC), J.P. Morgan ( JPM) and Goldman Sachs (GS). The sample period spans from 31 December 2015 to 29 December 2023, leading to $T=2013$ observations, and is available from https://finance.yahoo.com. In this illustration we set aside the last 500 observations to evaluate the out-of-sample predictive performance of the two GPU-Dirichlet models, 1) the NegBinC model (GPU Dirichlet with negative binomial generating function), and 2) the BernsteinCBP model  by \cite{burda14} ( GPU-Dirichlet with binomial generating function), to that of  four parametric models: three Archimedean copulas, rotated Gumbel, Clayton and rotated Joe, and the Gaussian copula.

\begin{figure}
\begin{center}\includegraphics[scale=0.7]{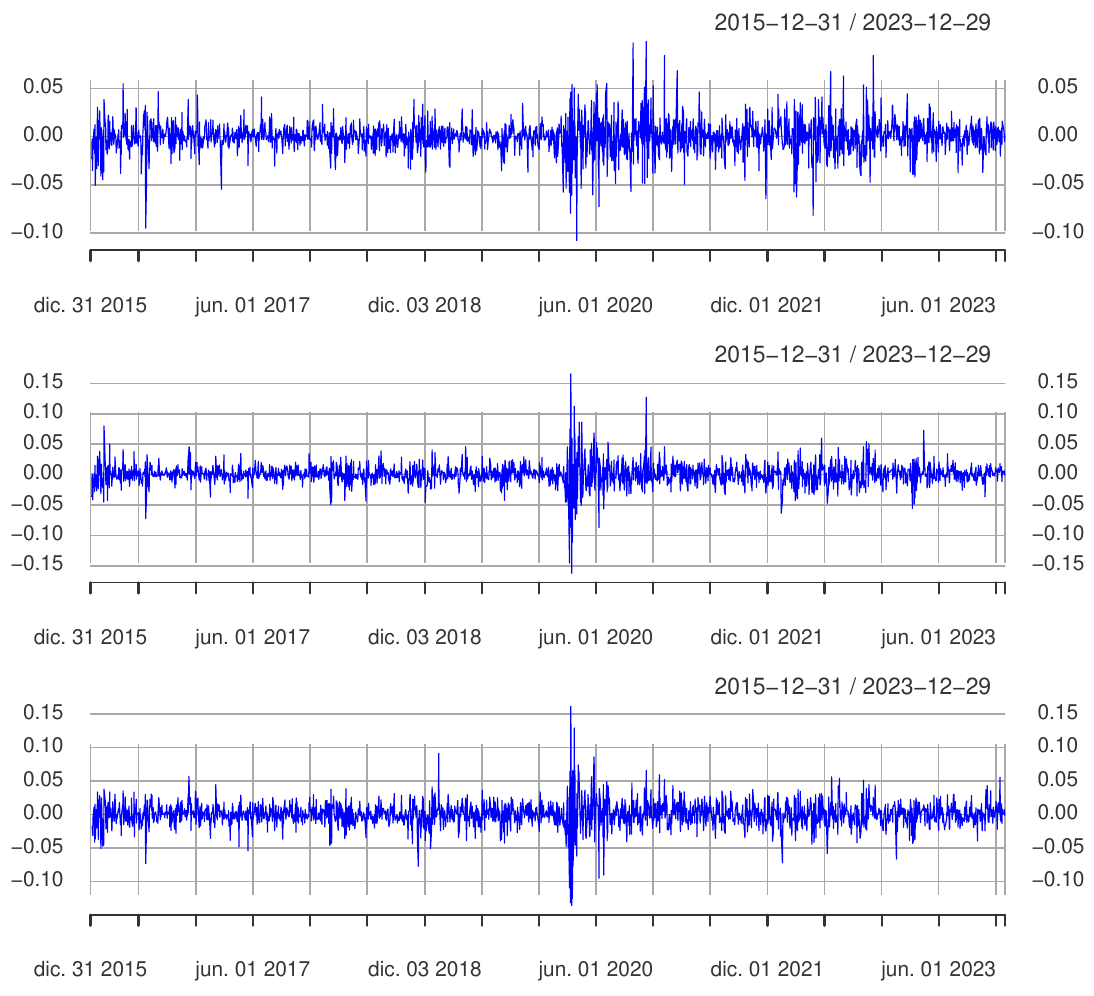}\end{center}
\vspace{-0.1in}

\caption{\small Time-series of the compounded returns of the HSBC (top), JPM (middle), GS (bottom).}\label{series} 
\end{figure}

Figure \ref{series} displays the time series of the compounded returns of HSBC (top), JPM (middle) and GS (bottom).  
The returns of the three banking stocks display slight asymmetry around the mean and excess kurtosis. Volatility clustering is also present with the Covid-19 lockdown period causing the highest volatility cluster. \newpage 

We therefore assume a nonparametric copula-GARCH model to capture joint dependence. We first fit a univariate AR(1)-EGARCH(1,1) model with the skewed Student-t innovations of \cite{steel98} to each series,\begin{eqnarray*}
	r_{it}&=& \mu_i+\phi_i r_{i,t-1}+\epsilon_{it}\sqrt{h_{it}}\\
	\log h_{it} &= &\omega_{i} +\alpha_i \epsilon_{i,t-1} + \gamma_i(\left| \epsilon_{i,t-1}\right| -E(\left| \epsilon_{i,t-1}\right|)) + \beta_i \log h_{i,t-1}
\end{eqnarray*} where $\epsilon_{it}$ follows a skewed Student-t with asymmetry and shape parameters,  $\xi_i$ and $\nu_i$, respectively.  We check the goodness of fit of the skewed Student-t choice by looking at the probability integral transform of the standardized residuals which approximately  follow Uniform(0,1) distributions in all cases.  Table \ref{EGARCH} displays the estimated coefficients for the marginal parameters. We then use the normalised ranked residuals to obtain the pseudo-data which are necessary to construct the copulas.
\begin{table}
	\begin{center}
		\begin{tabular}{ |l|c| c| c| c|c| c|c|c| } 
			\hline
			&  $\mu_i$& $\phi_i$&  $\omega_i$& $\alpha_i$  & $\beta_i$ &$\gamma_i$&  $\xi_i$& $\nu_i$ \\
			\hline
			HSBC &$-1.72 \ 10^{-5}$ & $-0.0305$& $-0.0839$ & $-0.0361$ & $0.9901$& $0.1335$ & $0.9542$ & $3.6510$\\
			JPM & $3.71 10^{-4}$& $-0.0055$ & $-0.2791$& $-0.0731$ & $0.9660$ & $0.1925$&$0.9841$ & $5.6065$\\
			 GS & $4.84 10^{-4}$& $-0.0234$&$-0.2513$ & $-0.1243$ & $0.9703$& $0.2165$& $1.0187$ & $4.7647$ \\
			 \hline 		
		\end{tabular}
	\end{center}
	\caption{ \small Estimated parameters of marginal AR(1)-EGARCH(1,1) models with the skewed Student-t innovations. }
	\label{EGARCH}
\end{table}

We apply the two GPU-Dirichlet models, NegBinC and BernsteinCBP, the three Archimedean copulas, rotated Gumbel, Clayton and rotated Joe, and the Gaussian copula to the pseudo-data of each pair of stock returns. Table \ref{LPSreturns} displays the LPS values for these six models. The LPS of the best model for each pair are in bold. It is clear the NegBinC model outperforms all other model for all pairs of stocks returns. 

\begin{table}
	\begin{center}
		\begin{tabular}{|l |c |c|c|c|c|c|c| } 
			\hline
			&	NegBinC &  BernsteinCBP  & RotGumbel & Clayton  & RotJoe & Gaussian \\
			\hline
			HSBC \& JPM &{\bf 0.1569} &  0.1454  & 0.1445  & 0.0767 & 0.0622& 0.1435 \\
			\hline 		
			
		HSBC \& GS  &{\bf 0.1208 } &  0.1109 & 0.1109 &0.0449 &  0.0363 & 0.1026 \\
			\hline 	
			JPM  \& GS   &{\bf 0.3285} &  0.3261 & 0.3107  & 0.0752& 0.0740&0.3260  \\
			\hline 				
		\end{tabular}
	\end{center}
	\caption{ \small LPS values for financial returns for the two Bayesian nonparametric copulas and some of parametric copulas.}
	\label{LPSreturns}
\end{table}

\section{Conclusion}\label{S7}
In this paper we have introduced a new family of Bayesian nonparametric copulas which we refer to as random GPU (RGPU) copulas, based on generalised partitions of unity under a Dirichlet process prior. We show that the RGPU copulas are rich class of models with the Bernstein copula model of \cite{burda14} (BernsteinCBP model) being a special case. We find that the RGPU copula we referred to as NegBinC accommodates both asymmetry in the dependence structure and tail dependence. Using the stick-breaking representation of the RGPUs, we have developed an MCMC algorithm to sample from the posterior distribution. Our simulated examples demonstrate that the NegBinC model outperforms the BernsteinCBP model of \cite{burda14} in capturing tail dependence. This advantage holds even in the presence of heterogeneity, suggesting that the NegBinC model adequately captures the dependence structure in both the tail and the centre of the distribution. We further illustrate the superior performance of  NegBinC model using real data from insurance and finance. 

GPU-copulas can capture single tail dependence, either upper or lower, \citep[see][]{pfe16, pfe17}. For our future work on RGPUs we will explore the development of theoretical two-tailed RGPU copulas. As discussed in Section \ref{S4}, our MCMC scheme is scalable to high-dimensional settings, and we intend to extend the RGPUs to higher dimensions, as the computational time does not increase significantly with the dimension. The assumption of a uniform smoothing parameter, $\theta$, across all margins can be overly restrictive, especially in practical applications, as it implies identical tail dependence for all pairs of variables. To address this, we are currently working on extending the RGPU to model, to allow for different smoothing parameters for each margin and incorporate a time-varying nonparametric framework, that is flexible and able to more accurately capture the dependence in financial time series.


\section*{Acknowledgments}

The first author  is supported by the grant PID2022-138114NB-I00/AEI/10.13039/501100011033 (\textit{High dimensional dependence modeling}) from the Spanish State Research Agency (Ministerio de Ciencia e Innovaci\'on).

\bibliographystyle{ba}
\bibliography{copulas}

\end{document}